\documentclass[letterpaper, 10 pt, conference]{ieeeconf}  

\IEEEoverridecommandlockouts                              

\usepackage{tikz}
\usetikzlibrary{positioning, shapes, arrows}
\usepackage{circuitikz}
\usepackage{graphicx}
\usepackage{array}
\usepackage{amsfonts}
\usepackage{stfloats}
\usepackage[noadjust]{cite}
\usepackage{booktabs}
\usepackage{multirow}
\usepackage{multicol}
\usepackage{amsmath}
\usepackage{amssymb}
\usepackage[noadjust]{cite}
\usepackage{pdfpages}

\newtheorem{thm}{Theorem}

\newtheorem{rem}{Remark}
\newtheorem{sass}{Standing Assumption}
\newtheorem{prop}{Proposition}

\DeclareMathOperator*{\argmin}{arg\,min}
\DeclareMathOperator*{\argmax}{arg\,max}
\makeatletter
\newcommand*{\rom}[1]{\expandafter\@slowromancap\romannumeral #1@}
\makeatother

\title{\LARGE \bf
Hybrid low-dimensional limiting state of charge estimator for multi-cell lithium-ion batteries
}

\author{Mira Khalil$^{1,2}$, 
        Romain Postoyan$^{1}$,
        St{\' e}phane Ra{\"e}l$^{2}$ and 
        Dragan Ne{\v{s}}i{\'c}$^{3}$
\thanks{*This work was supported by the French PIA project
“Lorraine Universit{\' e} d’Excellence”, reference ANR-15-IDEX-04-LUE.}
\thanks{$^{1}$Universit{\' e} de Lorraine, CNRS, CRAN, F-54000 Nancy, France.
{\tt(mira.khalil@univ-lorraine.fr,}
{\tt romain.postoyan@univ-lorraine.fr).}}
\thanks{$^{2}$Universit{\' e} de Lorraine, GREEN, F-54000 Nancy, France.
{\tt(stephane.rael@univ-lorraine.fr).}}
\thanks{$^{3}$Department of Electrical and Electronic Engineering, The University of Melbourne, Parkville, 3010 Victoria, Australia.
{\tt(dnesic@unimelb.edu.au).}}
}

\begin{document}

\maketitle
\thispagestyle{empty}
\pagestyle{empty}

\begin{abstract}
The state of charge (SOC) of lithium-ion batteries needs to be accurately estimated for safety and reliability purposes. For battery packs made of a large number of cells, it is not always feasible to design one SOC estimator per cell due to limited computational resources. Instead, only the minimum and the maximum SOC need to be estimated. The challenge is that the cells having minimum and maximum SOC typically change over time. In this context, we present a low-dimensional hybrid estimator of the minimum (maximum) SOC, whose convergence is analytically guaranteed. We consider for this purpose a battery consisting of cells interconnected in series, which we model by electric equivalent circuit models. We then present the hybrid estimator, which runs an observer designed for a single cell at any time instant, selected by a switching-like logic mechanism. We establish a practical exponential stability property for the estimation error on the minimum (maximum) SOC thereby guaranteeing the ability of the hybrid scheme to generate accurate estimates of the minimum (maximum) SOC. The analysis relies on non-smooth hybrid Lyapunov techniques. A numerical illustration is provided to showcase the relevance of the proposed approach.

\end{abstract}

\section{Introduction}
\label{intro}
Lithium-ion batteries offer many advantages over other energy storage technologies in terms of  weight, volume capacity, power density and absence of memory effect. However, they also require a battery management system (BMS) for safety and reliability purposes. The challenge is that only the battery voltage, its current and possibly its temperature are typically measured by sensors while the BMS also needs information about the battery internal states, in particular the state of charge (SOC). In this context, an abundant literature on the SOC estimation is available, see, e.g., \cite{HANNAN_2017_RSER,Barillas_2015_AppliedEnergy} and the references therein. A common approach consists in designing an observer based on a mathematical model of the battery internal dynamics e.g., \cite{Moura_2017_TCST,DreefDonkers_CDC_2018}. The vast majority of the related techniques focuses on single cell batteries. It appears that to meet the voltage and power requirements of some applications such as electric vehicles, hundreds of lithium-ion cells are interconnected in series, parallel or series-parallel to form a \textit{battery pack}. In this setting, we may not be able to design one observer per cell, as this would require significant computational capacity from the BMS. This motivates the design of low-dimensional estimation algorithms for multi-cell batteries, see e.g., \cite{Plett_2009,HUA_2015_JPS,Mawonou_2018_IECON,Liu_2010_CSEDM,ZHOU_2019_JPS,Moura_ACC_2020}. An important problem in this context is to estimate the states of the cells which respectively have the minimum SOC during discharge and maximum SOC during charge at a given time, we talk of \textit{limiting cells}. Indeed, monitoring the states of the limiting cells ensures that all the cells in the pack are within the operating limits and hence prevents safety hazards like over-charging and over-discharging. One of the challenges of this approach is that the limiting cells typically change over time because of the cells heterogeneity, which is due to the production process and the operating conditions. Several techniques have been developed to address this challenge, see e.g. \cite{Liu_2010_CSEDM,ZHOU_2019_JPS,Mawonou_2018_IECON}, but with no analytical guarantees as far as we know.  

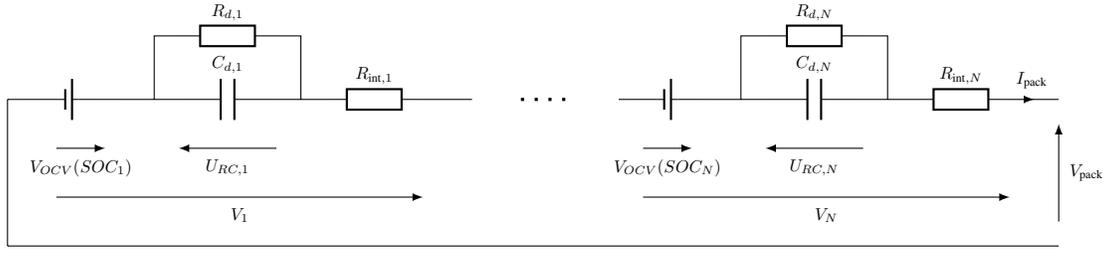
\begin{figure*}[t]
\begin{center}
\begin{circuitikz}[scale=0.65, transform shape] \draw
 (0,6) to[battery1] (-2.5,6)
(0,6)
  to[C=$C_{d,1}$] (4,6) to[european resistor=$R_{\text{int},1}$] (6,6) -- (7,6) 
 
(19,3)--(-2.5,3)
(0.5,6) --  (0.5,7.3)to[european resistor=$R_{d,1}$] (3.5,7.3) -- (3.5,6) 
(-2.5,6)--(-2.5,3)

(12,6) to[battery1] (10,6)
(12,6)
  to[C=$C_{d,N}$] (16,6) to[european resistor=$R_{\text{int},N}$] (18,6) -- (19,6) 

(12.5,6) --  (12.5,7.3)to[european resistor=$R_{d,N}$] (15.5,7.3) -- (15.5,6) 
;
\draw[loosely  dotted,very thick] (8,6) -- (9,6);
\draw[-latex] (-1.5,5) -- node[below=1mm] {$V_{OCV}(SOC_1)$} (-0.5, 5);
\draw[-latex] (19,3.5) -- node[right=1mm] {$V_{\text{pack}}$} (19, 5.5);
\draw[-latex] (3,5) -- node[below=1mm] {$U_{RC,1}$} (1, 5);
\draw[-latex] (18.4,6) -- node[above=1mm] {$I_{\text{pack}}$} (18.5, 6);
\draw[-latex] (10.5,5) -- node[below=1mm] {$V_{OCV}(SOC_N)$} (11.5, 5);
\draw[-latex] (15,5) -- node[below=1mm] {$U_{RC,N}$} (13, 5);
\draw[-latex] (10.5,4) -- node[below=1mm] {$V_N$} (18, 4);
\draw[-latex] (-1.5,4) -- node[below=1mm] {$V_1$} (6, 4);
\end{circuitikz}    
\end{center}
\caption{Schematic diagram for the lithium-ion battery pack consisting of $N$ cells in series modeled by a first order ECM.}
\label{Figure_ECM_Battery_Pack}
\end{figure*}

In this work, we present a low-dimensional hybrid model that estimates the state of the cell having at a given time the minimum SOC, whose convergence is analytically guaranteed. 
We focus on the cell with the minimum SOC without loss of generality as the presented results apply \textit{mutatis mutandis} for the estimation of the state of the cell with the maximum SOC. We consider for this purpose a series interconnection of $N$ cells, with $N \in \mathbb{Z}_{>0}$, see Figure \ref{Figure_ECM_Battery_Pack}. We model each cell by a first order equivalent circuit model (ECM) as it provides a good trade-off between complexity and accuracy. We allow the parameters of each cell model to differ. We then present the hybrid estimator. At any continuous time, we run one nonlinear observer for the cell determined by a selection variable, which we design. When this selection variable changes its value, a jump occurs, the estimated SOC exhibits a jump and the observer is then run for the newly selected cell. This selection variable relies on an on-line estimate of the open circuit voltages (OCV) of each cell. This is a key difference with \cite{Liu_2010_CSEDM}, which considers the cells voltage and with \cite{ZHOU_2019_JPS,Mawonou_2018_IECON}, which ignore the impact of $U_{RC,i}$, that is the voltage across the parallel branch representing the diffusion phenomena within the cells, and consisting of a capacity and a resistance (R-C), for all $i \in \{1,\hdots,N\}$, see Figure \ref{Figure_ECM_Battery_Pack}.
We model the overall system as a hybrid system using the formalism of \cite{Sanfelice_HSBook} and its extension to allow for continuous-time inputs in \cite{Heemels_CDC_2021}. The main result is a  practical exponential stability property for the minimum SOC estimation error. The stability results are established by constructing a new, non-smooth hybrid Lyapunov function. We also proved that the proposed estimation scheme does not generate Zeno solutions for bounded input currents, which is always the case in practice. The implementation of the proposed estimation scheme only requires $2$ variables, if we ignore the selection variable, which is a logic variable, constant between two successive jumps, while a brute force approach consisting of designing one observer per cell would require at least a $2N$-dimensional observer and even a $6N$-dimensional observer if we would be running a Kalman-like filter because of the covariance matrices. Simulation results finally illustrate the relevance of the proposed approach on a battery pack made of $200$ cells interconnected in series.

The rest of this paper is organised as follows. In Section \ref{Notation}, we introduce some notations. The battery pack model is given in Section \ref{BatteryModel}. The hybrid estimator is presented in Section \ref{HybridDesign}. The main stability results are provided in Section \ref{StabilityGuar}. Numerical simulations are given in Section \ref{Simu}. Section \ref{conclu} concludes the paper.

\section{Notation and preliminaries}
\label{Notation}

Let $\mathbb{R}$ be the set of real numbers, $\mathbb{R}_{>0}:=(0,\infty)$, $\mathbb{R}_{\geq0}:=[0,\infty)$, $\mathbb{Z}$ be
the set of integers, $\mathbb{Z}_{>0}:=\{1,2,3,...\}$, $\mathbb{Z}_{\geq 0}:=\{0,1,2,3,...\}$. Given square matrices $A_1,..., A_n$, diag($A_1,..., A_n$) is the block diagonal matrix, whose block diagonal components are $A_1,..., A_n$. Given a vector $x \in \mathbb{R}^n$, $x^\top$ denotes the transpose of $x$. Given $x\in \mathbb{R}^{n}$ and $y\in \mathbb{R}^{m}$ with $n,m\in\mathbb{Z}_{>0}$, we use the notation $(x,y)$ for $(x^\top,y^\top)^\top$. For a vector $x \in \mathbb{R}^N$, $|x|$ stands for its Euclidean norm. Given $f:\mathbb{R}_{\geq 0}\to\mathbb{R}^n$ with $n\in\mathbb{Z}_{>0}$, we write $f\in\mathcal{L}_\infty$ if $\underset{{\tau\geq 0}}{\text{ess.sup}} \ |f(\tau)|<\infty$. 

We will model the proposed estimator together with the plant as a hybrid system with continuous-time inputs of the form \cite{Heemels_CDC_2021,Sanfelice_HSBook}
\begin{equation}
    \begin{aligned}
    \dot q &= f(q,u) \quad  (q,u)\in \mathcal{C}\\
    q^+ &\in G(q) \quad \quad    (q,u)\in \mathcal{D},
\end{aligned}
\end{equation}
where $\mathcal{C} \subseteq \mathbb{R}^{n_q}$ is the flow set, $\mathcal{D} \subseteq \mathbb{R}^{n_q}$ is the jump set, $f$ is the flow map and $G$ is the set-valued jump map. We adopt the notion of solutions in \cite[Definition 4 (S1-e)(S2)]{Heemels_CDC_2021} considering $u$ as a c\`adl\`ag (“continue à droite, limite à gauche”) input \cite[Definition 8]{Heemels_CDC_2021}. We consider hybrid
time domains and hybrid arcs as defined in \cite{Sanfelice_HSBook}. The notation $(t,j) \geq (t',j')$ means that $t \geq t'$ and $j \geq j'$, where $(t,j), (t',j')\in \mathbb{R}_{\geq 0}\times \mathbb{Z}_{\geq 0}$.
Let $w$ be a hybrid arc, we define $\|w\|_{(t,j)} :=
\max \{ \underset{{(t',j' )\in \text{dom $w$} \backslash \Gamma(w), (0,0)\leq(t',j' )\leq(t,j)}}{\text{ess.sup}} |w(t', j')|, \\ \underset{(t',j' )\in \Gamma(w), (0,0)\leq(t',j' )\leq(t,j)}{\text{sup}}
|w(t', j')|\}$ where $\Gamma(w)$ is
the set of all $(t', j') \in$ dom $w$ such that $(t', j' + 1) \in$ dom $w$. Finally, we use $V^{\circ}(x;v):=\lim \text{sup}_{h \rightarrow 0^+, y \rightarrow x}   \frac{V(y+hv)-V(y)}{h}$ to denote
the Clarke generalized directional derivative of a Lipschitz
function $V$ at $x$ in the direction $v$ \cite{clarkeBook}.


\section{Battery pack modeling}
\label{BatteryModel}
We consider a battery pack consisting of $N$ lithium-ion cells in series, with $N \in \mathbb{Z}_{>0}$, where each cell is modeled by an ECM, see Figure \ref{Figure_ECM_Battery_Pack}. We first present the equations that describe the ECM of an individual cell. Then, we present a model of the battery pack.

\subsection{Individual cell}
Given $i \in \{1,\hdots,N\}$, the ECM of cell $i$ is given by
\begin{equation}
\begin{aligned}
   \frac{d U_{RC,i}}{d t}&= -\frac{1}{\tau_{d,i} } U_{RC,i} + \frac{1}{C_{d,i} } I_i
   \\
     \frac{d SOC_i}{d t}&= -\frac{1}{3600Q_i}I_i
\\ V_i&=-U_{RC,i}-R_{\text{int},i}I_i+V_{OCV}(SOC_i),
 \end{aligned}
 \label{ECMcelli}
\end{equation}
where $U_{RC,i} \in \mathbb{R}$ is the voltage across the parallel R-C branch, $\tau_{d,i}:=R_{d,i}C_{d,i}$ is the diffusion time constant, $R_{d,i} \in \mathbb{R}_{>0}$ is the diffusion resistance, $C_{d,i} \in \mathbb{R}_{>0}$ is the capacitance associated with the diffusion phenomena, $V_{OCV}(SOC_i) \in \mathbb{R}$ is the OCV related to $SOC_i \in [0,1]$ (or $[0\%,100\%]$) the state of charge of the cell, $Q_i \in \mathbb{R}_{>0}$ is the cell nominal capacity, $R_{\text{int},i} \in \mathbb{R}_{>0}$ is the ohmic resistance, $V_i \in \mathbb{R}$ is the cell voltage and $I_i$ is the cell current. Current $I_i$ is the input to the system in (\ref{ECMcelli}), which we know, and $V_i$ is the output, which we measure. Indeed, it is essential to monitor the output voltage of \emph{each} cell in  practice for safety purposes.
We make the next assumption on the cells parameters.
\begin{sass}[SA1]
 For any $i\in \{1,\hdots,N\}$, $\tau_{d,i}$ is constant and the parameters $R_{d,i}$, $R_{\text{int},i}$, $Q_i$ are constant and known.
 $\hfill \Box$
    \label{Ass_on_Parameters}
\end{sass}

 SA\ref{Ass_on_Parameters} allows the parameters of each cell model to differ, which is in line with practice where cells may differ because of the operating conditions and the production processes. SA\ref{Ass_on_Parameters} also states that the parameters in (\ref{ECMcelli}) are constant. When this is not the case but the parameters variations are ``small'', their impact will be negligible on the proposed design due to its intrinsic robustness properties. We plan to investigate the case where some of these parameters (significantly) vary with time in future work.

We make the next assumption on $V_{OCV}$.
\begin{sass}[SA2]
    All the cells have the same function $V_{OCV}$ and the following holds.
    \begin{itemize}
        \item[(i)] $V_{OCV}$ is continuously differentiable on $\mathbb{R}$.
        \item[(ii)] $V_{OCV}$ has strictly positive bounded minimum and maximum derivatives, i.e., $0<a_1:=\underset{z \in \mathbb{R}}{\min} \frac{\partial V_{OCV} }{\partial z}(z) \leq   \underset{z \in \mathbb{R}}{\max} \frac{\partial V_{OCV} }{\partial z}(z):=a_2<\infty$.
    \end{itemize} 
    $\hfill \Box$
    \label{Ass_on_Vocv}
\end{sass}

SA\ref{Ass_on_Vocv} is not restrictive for the following reasons.
The cells in a battery pack are generally of the same chemistry and from the same production batch. Hence, their OCV-SOC map are very similar and it is reasonable to assume the same $V_{OCV}$ for all the cells. Regarding item (i) of SA\ref{Ass_on_Vocv},
$V_{OCV}$ is generally given by a look-up table over $[0,1]$ $([0\%,100\%])$, which we can very well interpolate using a continuously differentiable function on $[0,1]$ $([0\%,100\%])$. We can then extrapolate the definition of $ V_{OCV}$ over $\mathbb{R}$ while preserving this regularity property so that item (i) holds. On the other hand, item (ii) is reasonable for most lithium-ion technologies such as lithium nickel manganese cobalt oxide (NMC) and lithium nickel cobalt aluminum oxide (NCA), more generally, for any battery technology with non-flat OCVs. Here as well, the strict monotonicity property of $V_{OCV}$ on $[0,1]$ $([0\%,100\%])$ can be extended to $\mathbb{R}$.


To derive a state space representation of cell $i$, we introduce the state vector $x_i:=(U_{RC,i},SOC_i) \in \mathbb{R}^2$, the input $u := I_\text{i} \in\mathbb{R}$ and the output $y_i:=V_i \in \mathbb{R}$. From (\ref{ECMcelli}), we obtain
\begin{equation}
\begin{cases}
    \Dot{x}_i=A_i x_i+ B_i u \\
    y_i=C x+D_i u+V_{OCV}(SOC_i),
\end{cases}
\label{Singlecell_SS}
\end{equation}
with $A_i:=
\begin{pmatrix}
-\frac{1}{\tau_{d,i} } & 0\\
0 & 0
\end{pmatrix},  
B_i:=
\begin{pmatrix}
\frac{1}{C_{d,i}}\\
-\frac{1}{3600Q_i}
\end{pmatrix}$, $D_i:=-R_{\text{int},i}$ for $i \in \{1,\hdots,N\}$ and $C:=\begin{pmatrix}
    -1 & 0
\end{pmatrix}$.
\begin{rem}
In addition to its simplicity compared to other battery models, the cell ECM presents an explicit nonlinear relationship between the OCV and the SOC as seen in the output equation of (\ref{Singlecell_SS}). In Section \ref{HybridDesign}, we exploit this property to select the cell having potentially the minimum SOC. 
$\hfill \square$
    \label{Rem_on_ECM}
\end{rem}
\subsection{Multi-cell}
In Figure \ref{Figure_ECM_Battery_Pack}, $N$ single cell first order ECMs are interconnected in series, where $V_{\text{pack}} \in \mathbb{R}$ is the measured voltage of the lithium-ion battery pack and $I_{\text{pack}} \in \mathbb{R}$ is its current. We note that in Figure \ref{Figure_ECM_Battery_Pack} the sum of all the $V_i$ is equal to $V_{\text{pack}}$ and all the $I_i$ are equal and denoted by $I_{\text{pack}}$. 

We define the state vector $x:=(x_1,x_2, \hdots, x_N) \in\mathbb{R}^{2N}$, the input $u := I_\text{pack} \in\mathbb{R}$  and the output $y := (y_1, \hdots, y_N) \in\mathbb{R}^N$ as well as $SOC:=(SOC_1,\hdots,SOC_N)$ and $U_{RC}:=(U_{RC,1},\hdots,U_{RC,N}) \in \mathbb{R}^{N}$. Given the series interconnection in Figure \ref{Figure_ECM_Battery_Pack} and in view of (\ref{Singlecell_SS}), we derive
\begin{equation}
\begin{cases}
    \Dot{x}=\textbf{A} x + \textbf{B} u \\
    y=\textbf{C} x+\textbf{D} u+V_{OCV}(SOC),
\end{cases}
\label{Multicell_SS}
\end{equation}
where $\textbf{A}:=\text{diag}(A_1, A_2, \hdots,A_N) \in \mathbb{R}^{2N \times 2N}$, $\textbf{B}:=(B_1,B_2\hdots,B_N) \in \mathbb{R}^{2N \times 1}$, $\textbf{C}:=\text{diag}(C, \hdots, C) \in \mathbb{R}^{N \times 2N}$, $\textbf{D}:=(D_1, D_2, \hdots,D_N) \in \mathbb{R}^{N \times 1}$ and $V_{OCV}(SOC):=(V_{OCV}(SOC_1),\hdots, V_{OCV}(SOC_N)) \in \mathbb{R}^{N}$.

\subsection{Limiting cell}
Our objective is to estimate on-line the minimum SOC of system (\ref{Multicell_SS}), while not running an observer for each cell to save computational resources. The estimation of the minimum SOC is essential in discharge to prevent going beyond the lower operating limits and hence over-discharging. In charge, the estimation of the maximum SOC is needed, which we do not explicitly address in this work as the results apply similarly. At any given time, the minimum SOC is formally defined as 
\begin{equation}
    \begin{aligned}
        SOC_{\min}&:=\underset{i \in \{1,\hdots,N\}}{\min} SOC_i. \\
    \end{aligned}
    \label{Def_LimitingSOC}
\end{equation}
In the case where we have multiple cells having minimum SOC at a given time, we consider any of them to be the cell with minimum SOC.
In the next section, we present a hybrid model that estimates $SOC_{\min}$ by running an observer for a single cell at any time instant. 

\section{Hybrid design}
\label{HybridDesign}

We present in this section the low-dimensional hybrid design, whose purpose is to estimate $SOC_{\min}$ as defined in (4).
We first present the continuous-time dynamics of the hybrid estimator in Section \ref{HybridDesign_obs} then its discrete-time dynamics in Section \ref{HybridDesign_cellselect}. Section \ref{ Hybriddesign_FlowandJumpsets} explains the switching logic. Finally, the overall system is modeled as a hybrid system of the form \cite{Sanfelice_HSBook,Heemels_CDC_2021}, where a jump corresponds to a change of the selected cell in Section \ref{HybridDesign_HM}.

\subsection{Continuous-time dynamics}
\label{HybridDesign_obs}
We denote by $\sigma$ the logic variable, which is used to select one cell at any given time. Hence, $\sigma$ takes values in $\{1,\hdots,N\}$ and is constant between two successive jumps, i.e.,
\begin{equation}
\dot \sigma = 0.
\label{sigma_CT}
\end{equation}
The estimates for $U_{RC,\sigma}$ and $SOC_\sigma$ are denoted $\widehat{U}_{RC,\sigma}$ and $\widehat{SOC}$, respectively, and are given by, between two successive jump times, 
\begin{equation}
    \begin{aligned}
        \dot{{\overline{U}}}_{RC}&= -\frac{1}{\tau_d}{\overline{U}}_{RC}+u \\
         \dot{\widehat{{SOC}}}&= -\frac{1}{3600Q_{ \sigma}}u + \ell(y_{ \sigma}-\hat{ {y}}) \\
        \widehat{U}_{RC,\sigma}&= \tfrac{1}{C_{d,\sigma}}\overline{U}_{RC} \\
         \hat{ {y}}&= -\widehat{U}_{RC, \sigma} + D_{ \sigma} u + V_{OCV}(\widehat{{SOC}}),
    \end{aligned}
    \label{SOCminObs}
\end{equation}  
where ${{\overline{U}}}_{RC}$ is an intermediate variable we use to generate $\widehat U_{RC,\sigma}$. This choice is motivated by the fact that by running $\overline U_{RC}$, we can also generate estimates of $U_{RC,i}$ for any $i\in\{1,\hdots,N\}$ and not only for $i=\sigma$, which will play a key role in the design of the discrete-time dynamics of selection variable $\sigma$ and of the switching logic, see Sections \ref{HybridDesign_cellselect} and \ref{ Hybriddesign_FlowandJumpsets}. Parameter $\tau_d$ can in principle take any value in $\mathbb{R}_{>0}$. When $\tau_{d,i}$ is known for any $i \in \{1,\hdots,N\}$, we can define it as $\tau_d:=\tfrac{1}{N}\sum_{i=1}^N \tau_{d,i}$, that is the average diffusion time constant, or by $\tau_d \in \underset{a>0}{\argmin}{\underset{i \in \{1,\hdots,N\}}{\max}|\tfrac{1}{a}-\tfrac{1}{\tau_{d,i}}|}$ as the mismatch between $\tfrac{1}{\tau_d}$ and $\tfrac{1}{\tau_{d,i}}$ has an impact on the accuracy of the estimates as shown in Section \ref{StabilityGuar}. In the case where the $\tau_{d,i}$'s are not known but given by a distribution with a well-defined mean, we can take $\tau_d$ to be the mean of the distribution. On the other hand, the dynamics of $\widehat{SOC}$ is made of a copy of the dynamics of $SOC_{\sigma}$ with an additive correction term $\ell(y_\sigma-\hat y)$ where $\hat{ {y}}$ is the estimate of $y_{ \sigma}$ and $\ell$ is a design parameter that can take any value in $\mathbb{R}_{>0}$.


\subsection{Discrete-time dynamics}
\label{HybridDesign_cellselect}
The role of variable $\sigma$ is to select the cell corresponding to the minimum SOC at any time instant. Of course, we cannot directly compare $SOC_i$, for all $i\in\{1,\ldots, N\}$, for this purpose since $SOC_i$ is not measured. We can also not rely on estimates of $SOC_i$, for any $i\in\{1,\hdots,N\}$, as only $SOC_\sigma$ is estimated in view of (\ref{SOCminObs}). The idea is to exploit estimates of $V_{OCV}(SOC_i)$ for any $i\in\{1,\hdots,N\}$ as $V_{OCV}$ is a monotone function of the SOC by item (ii) of SA\ref{Ass_on_Vocv}. Hence, if our estimates of $V_{OCV}(SOC_i)$ are accurate, taking the cell with the minimum estimated $V_{OCV}(SOC_i)$ would give us the cell with the minimum SOC.
From (\ref{ECMcelli}), we have for any cell $i \in \{1,\hdots,N\}$
\begin{equation}
    V_{OCV}(SOC_i)=V_i+U_{RC,i}+R_{\text{int},i}I_\text{pack}.
    \label{OCVeq}
\end{equation}
While $V_i$ and $R_{\text{int},i}I_\text{pack}$ are known, $U_{RC,i} $ is not. We therefore introduce $z_i$ as an estimate of $V_{OCV}(SOC_i)$
    \begin{equation}
    \begin{aligned}
        z_i&:=V_i+ {{\widehat{U}}}_{RC,i} +R_\text{int,i}I_\text{pack},
    \end{aligned}
    \label{zi}
\end{equation}
where ${{\widehat{U}}}_{RC,i}=\tfrac{1}{C_{d,i}}{{\overline{U}}}_{RC}$ is the estimate of $U_{RC,i}$. We note that the introduction of the intermediate variable ${\overline{U}}_{RC}$ in (\ref{SOCminObs}) avoids the need to estimate $U_{RC,i}$ of each cell and thus prevents from increasing the dimension of the estimator. 

Based on (\ref{zi}), $\sigma$ changes values as
\begin{equation}
    \begin{aligned}
        \sigma^+ &\in \underset{i \in \{1,\hdots,N\}\backslash\{{ \sigma}\}}{\argmin }z_i.
    \end{aligned}
    \label{sigmadef_ti}
\end{equation}
In the case where we have multiple cells having minimum $z_i$ at the same time, we select randomly one of them. Furthermore, at a switching time, we define
\begin{equation}
    \begin{aligned}
         \overline{U}_{RC}^+&=\overline{U}_{RC}\\
        \widehat{SOC}^+ &= {V_{OCV}}^{-1} (\underset{i \in \{1,\hdots,N\}\backslash\{{ \sigma}\}}{\min} z_i).
    \end{aligned}
    \label{discretetimedynamics_URCwidehatSOC}
\end{equation}
After a jump, $\overline{U}_{RC}$ remains the same while $\widehat{SOC}$ changes. The change we enforce on $\widehat{SOC}$ at jump times appears to be key to establish analytical guarantees on the convergence of the proposed hybrid estimator, see Section \ref{StabilityGuar}. We note that ${V_{OCV}}^{-1}$ is well defined as $V_{OCV}$ is bijective by item (ii) of SA\ref{Ass_on_Vocv}.
\begin{rem}
   In \cite{Liu_2010_CSEDM}, the lowest $V_i$ is used to estimate the SOC of the battery pack. In \cite{ZHOU_2019_JPS,Mawonou_2018_IECON}, $V_i+R_{\text{int},i}I_\text{pack}$ is the criterion upon which the limiting cells were identified. In this work, we propose an alternative approach where we exploit more accurate estimates of $V_{OCV}(SOC_i)$ thanks to estimates of $U_{RC,i}$ to select the cell for which we estimate the SOC and we provide analytical guarantees for the proposed hybrid estimator in Section \ref{StabilityGuar}.
   $\hfill  \square$
\end{rem}

\begin{rem}
In view of (\ref{sigmadef_ti}), the cell with the lowest $z_i$, i.e., estimated $V_{OCV}(SOC_i)$, is the selected cell at each jump time based on which the SOC is estimated and potentially is the cell having minimum SOC. 
 If each cell were to have its own function of the SOC, namely $V_{OCV,i}$, we may no longer obtain the cell with the minimum SOC by considering the cell with the lowest $V_{OCV,i}(SOC_i) $. 
 $\hfill \square$
    \label{Rem_on_V_ocv}
\end{rem}
\subsection{Switching logic}
\label{ Hybriddesign_FlowandJumpsets}

 We are now ready to present the switching logic. As long as we have for all $i \in \{1,\hdots,N\}\backslash\{{ \sigma}\}$, $V_{OCV}(\widehat{SOC})-\varepsilon \leq z_i $, where $\varepsilon \in \mathbb{R}_{>0}$ is a regularization parameter to avoid Zeno behavior (see Section \ref{StabilityGuar_MR}), we do not need to select a new cell and thus cell $\sigma$ remains the same as in (\ref{sigma_CT}). 
On the other hand, when there exists $i \in \{1,\hdots,N\}\backslash\{{ \sigma}\}$ such that $V_{OCV}(\widehat{SOC}) - \varepsilon \leq  z_i \leq V_{OCV}(\widehat{SOC})-\mu \varepsilon $, where $\mu \in (0,1]$ is a design parameter, then cell $\sigma$ changes as in (\ref{sigmadef_ti}).
We thus define the flow set $\mathcal{C}$ and the jump set $\mathcal{D}$, given parameter $\varepsilon \in \mathbb{R}_{>0}$ and $\mu \in (0,1]$, as
\begin{equation}
\begin{aligned}
  \mathcal{C}:=\{&q \in \mathcal{Q}: \forall i \in \{1,\hdots,N\}\backslash\{{ \sigma}\}, \\&V_{OCV}(\widehat{SOC})-\varepsilon\leq z_i \} \\
  \mathcal{D}:=\{&q \in \mathcal{Q}: \exists i \in \{1,\hdots,N\}\backslash\{{ \sigma}\},  \\&V_{OCV}(\widehat{SOC}) - \varepsilon \leq  z_i  \leq V_{OCV}(\widehat{SOC})-\mu \varepsilon   \}.
\end{aligned}
\label{FlowJumpSets}
\end{equation}
The definitions of $\mathcal{C}$ and $\mathcal{D}$ in (\ref{FlowJumpSets}) restrict the initial condition on the hybrid estimator. This is not an issue as we can take the initial values of $\widehat{SOC}$ and $\overline{U}_{RC}$ such that the initial condition of $q$ always lies in $\mathcal{C}\cup \mathcal{D}$ since the restrictions imposed by $\mathcal{C}$ and $\mathcal{D}$ can be verified based on the available data. On the other hand, the role of parameter $\mu$ is to enlarge set $\mathcal{D}$. Indeed, our results apply for $\mu=1$, however in this case $\mathcal{D}$ is of Lebesgue measure $0$, which may be difficult to implement in practice. This potential issue is overcome by taking $\mu\in(0,1)$. We also note that $\mathcal{D}\subset \mathcal{C}$, which is not an issue for the implementation and the analysis of the hybrid scheme.

\subsection{Hybrid model}
\label{HybridDesign_HM}
The overall state of the hybrid model is defined as $q:=(x,{\overline{U}}_{RC}, \widehat{{SOC}},  \sigma) \in \mathcal{Q}:=\mathbb{R}^{2N}\times \mathbb{R} \times \mathbb{R} \times  \{1,\hdots,N\}$. By collecting (\ref{Multicell_SS}), (\ref{sigma_CT}), (\ref{SOCminObs}), (\ref{sigmadef_ti}), (\ref{discretetimedynamics_URCwidehatSOC}), we obtain the following hybrid system
\begin{equation}
\begin{aligned}
    \dot q &= f(q,u) \quad  (q,u)\in \mathcal{C}\\
    q^+ &\in G(q) \quad \quad    (q,u)\in \mathcal{D},
\end{aligned}
\label{compacthybridsystem}
\end{equation}
where $f(q,u):=(\textbf{A} x + \textbf{B} u, -\frac{1}{\tau_d}{\overline{U}}_{RC}+u , -\frac{1}{3600Q_{ \sigma}}u + \ell(y_{ \sigma}-\hat{ {y}}),0)$ with $y_\sigma=C x+D_\sigma u+V_{OCV}(SOC_\sigma)$, $ \hat{ {y}}=-{{\widehat{U}}}_{RC,\sigma} + D_{ \sigma} u + V_{OCV}(\widehat{{SOC}})$, $G(q):=(x,{\overline{U}}_{RC},{V_{OCV}}^{-1} (\underset{i \in \{1,\hdots,N\}\backslash\{{ \sigma}\}}{\min} z_i), \underset{i \in \{1,\hdots,N\}\setminus\{{ \sigma}\}}{{\argmin}} z_i)$ with $z_i=V_i-D_iu+ {{\widehat{U}}}_{RC,i}$, $ {{\widehat{U}}}_{RC,i}=\tfrac{1}{C_{d,i}}{{\overline{U}}}_{RC}$ for any $q \in \mathcal{Q}$ and $u \in \mathbb{R}$.

While the dimension of the overall hybrid system is $2N+2$, the hybrid estimator to be implemented is only of dimension $2$. Indeed, if we disregard $\sigma$, which is a scalar logic variable constant on flow, the observer, which needs to be run in practice is given by (\ref{SOCminObs}), (\ref{discretetimedynamics_URCwidehatSOC}) and is of dimension $2$. 
\begin{rem}
To estimate the maximum SOC instead of the minimum one, the hybrid estimator needs to be modified as follows. First, we take $\sigma^+ \in \underset{i \in \{1,\hdots,N\}\backslash\{\sigma\}}{\argmax} z_i$ and $\widehat{SOC}^+ = V_{OCV}^{-1}(\underset{i \in \{1,\hdots,N\}\backslash\{{ \sigma}\}}{\max} z_i)$. Second, the set $\mathcal{C}$ needs to be defined as $\{q \in \mathcal{Q}: \forall i \in \{1,\hdots,N\}\backslash\{\sigma\}$, $V_{OCV}(\widehat{SOC})  \geq z_i-\varepsilon\}$ and the set $\mathcal{D}$ as $\{q \in \mathcal{Q}: \exists i \in \{1,\hdots,N\}\backslash\{\sigma\}$, $z_i-\varepsilon \leq V_{OCV}(\widehat{SOC}) \leq z_i -\mu \varepsilon\}$ still with $\mu\in(0,1]$. We plan to address in detail the case where both the minimum and the maximum SOC need to be estimated in a future work.
$\hfill \square$
    \label{Rem_MaximumSOCestimator}
\end{rem}

\section{Stability guarantees}
\label{StabilityGuar}
We establish in this section stability properties for the proposed hybrid estimator presented in Section \ref{HybridDesign}. We first state the main results in Section \ref{StabilityGuar_MR} and postpone their proofs to Section \ref{StabilityGuar_Proofs}.
\subsection{Main Results}
\label{StabilityGuar_MR}
We first state an input-to-state stability property for the estimation error on $U_{RC}$. 
\begin{prop}
Given any $\ell >0$ and $\varepsilon>0$, consider system (\ref{compacthybridsystem}), for any c\`adl\`ag input $u$, any corresponding solution $q$ to (\ref{compacthybridsystem}) satisfies, for all $(t,j) \in$ dom $q$,
   \begin{equation}
\begin{aligned}
     |U_{RC}(t,j)&-\widehat{U}_{RC}(t,j)|\\& \leq \sqrt{N} |U_{RC}(0,0)-\widehat{U}_{RC}(0,0)| e^{-\frac{1}{2\tau_d}t} \\& \quad + \sqrt{N}\tau_d  d \|U_{RC}\|_{(t,j)},
    \end{aligned}
    \label{StabilityProperty_delta_URC}
\end{equation}
where $\widehat{U}_{RC}:=(\widehat{U}_{RC,1},\hdots,\widehat{U}_{RC,N})$ and $d:=\underset{i \in \{1,\hdots,N\}}{\max}\big\{\big|\tfrac{1}{\tau_d}-\tfrac{1}{\tau_{d,i}} \big| \big\}$.
  $\hfill \Box$
\label{Th_StabilityProperty_delta_URC}
\end{prop}

Proposition \ref{Th_StabilityProperty_delta_URC} implies that system (\ref{compacthybridsystem}) satisfies an input-to-state stability property with respect to $U_{RC}-\widehat{U}_{RC}$. Inequality (\ref{StabilityProperty_delta_URC}) states that $|U_{RC}-\widehat{U}_{RC}|$ is upper-bounded along any solution to (\ref{compacthybridsystem}) by an exponentially decaying term involving the difference of the initial conditions of $U_{RC}$ and $\widehat{U}_{RC}$ and a term involving $\|U_{RC}\|_{(t,j)}$ times a constant related to the difference of the diffusion time constants $\tau_{d,i}$ with $\tau_d$. In the case where all the cells in the battery pack have the same $\tau_{d,i}$ and $\tau_d=\tau_{d,i}$, $d=0$ and we obtain an exponential stability property. We also note that as the number of cells $N$ increases, the overshoot of $U_{RC}-\widehat{U}_{RC}$ may increase.
Next, we show that system (\ref{compacthybridsystem}) also satisfies an input-to-state stability property with respect to $SOC_{\sigma}-\widehat{SOC}$, which is key to establish the main result of this section stated afterwards in Theorem \ref{TH_Stability_Property_SOCmin}.

\begin{prop}
Given any $\ell >0$ and $\varepsilon>0$, consider system (\ref{compacthybridsystem}), for any c\`adl\`ag input $u$, any corresponding solution $q$ to (\ref{compacthybridsystem}) satisfies, for all $(t,j) \in$ dom $q$,
\begin{equation}
 \begin{aligned}
         |SOC_{\sigma(t,j)}(t,j)-\widehat{SOC}(t,j)| \leq & c_1|e(0,0)|e^{-c_2t} \\&+ c_3 {\varepsilon}+c_4 d {\|U_{RC}\|_{(t,j)}},
    \end{aligned}
    \label{Th1_property}
\end{equation}
\label{TH_Stability_Property_e}
\end{prop}
where  $e:= (SOC_{ \sigma }-\widehat{SOC}, U_{RC,1}-\widehat{U}_{RC,1}, \hdots,$
$ U_{RC,N}-\widehat{U}_{RC,N})$, $c_1:=\sqrt{\max\{1,\frac{4}{a_1^2}\}}$, $c_2:=\tfrac{a}{2}$, $c_3:=\frac{4}{a_1}$, $c_4:=\frac{2}{a_1}\sqrt{ \frac{\tau_d }{a}}$, $a:={\min}\big\{\ell a_1,\tfrac{1}{\tau_d}\big\}$, $a_1$ and $a_2$ in item (ii) of SA\ref{Ass_on_Vocv}, $\varepsilon$ in (\ref{FlowJumpSets}) and $d$ in Proposition \ref{Th_StabilityProperty_delta_URC}.  $\hfill \Box$

Proposition \ref{TH_Stability_Property_e} guarantees a two-measure input-to-state stability property \cite{Cai_TAC_2007}. In particular, the norm of the mismatch between $SOC_\sigma$ and $\widehat{SOC}$ is shown to be upper-bounded, along any solution to (\ref{compacthybridsystem}), by an exponentially decaying term of the difference of the initial estimation errors and two additive terms one due to $\varepsilon$, which is expected in view the definitions of $\mathcal{C}$ and $\mathcal{D}$ in (\ref{FlowJumpSets}), and another involving $\|U_{RC}\|_{(t,j)}$ times $d$ consistently with Proposition \ref{Th_StabilityProperty_delta_URC}.
In (\ref{Th1_property}), we see that we can speed up the convergence to zero of the decaying term as much as we want by increasing $\ell$ and decreasing $\tau_d$, however in this case $d$ would grow, which would lead to a bigger ultimate error.
Because $SOC_{ \sigma}$ is not necessarily equal to $SOC_{\min}$, it remains to relate $\widehat{SOC}$ to $SOC_{\min}$: this is done in the next theorem.
\begin{thm}
Given any $\ell >0$ and $\varepsilon>0$, consider system (\ref{compacthybridsystem}), for any c\`adl\`ag input $u$, any corresponding solution $q$ to (\ref{compacthybridsystem}) satisfies, for all $(t,j) \in$ dom $q$,
\begin{equation}
\begin{aligned}
      |\widehat{{SOC}}&(t,j) -SOC_{\min}(t,j)|  \\&\leq \left(\tfrac{\sqrt{N}}{a_1}+c_1\right)|e(0,0)|e^{-bt} \\& \quad + \left(\tfrac{1}{a_1}+c_3\right)\varepsilon + \left(\tfrac{\sqrt{N}\tau_d }{a_1}+c_4\right) d \|U_{RC}\|_{(t,j)},
    \end{aligned}
    \label{StabilityProperty_with_SOCmin}
\end{equation}
where $b:=\min\{\tfrac{1}{2\tau_d},c_2\}$, $d$ is defined in Proposition \ref{Th_StabilityProperty_delta_URC} and $e$, $c_1$, $c_2$, $c_3$, $c_4$ are given in Proposition \ref{TH_Stability_Property_e}.
  $\hfill \Box$
    \label{TH_Stability_Property_SOCmin}
\end{thm}

Theorem \ref{TH_Stability_Property_SOCmin} guarantees that $\widehat{SOC}$ exponentially converges to $SOC_{\min}$ up to a tunable error due to $\varepsilon$ in (\ref{FlowJumpSets}) and an additional error due to the mismatch between $\tfrac{1}{\tau_d}$ and $\tfrac{1}{\tau_{d,i}}$, with $i\in\{1,\hdots,N\}$. Because the latter are typically small and $\|U_{RC}\|_{(t,j)}$ in (\ref{StabilityProperty_with_SOCmin}) is bounded in practice, (\ref{StabilityProperty_with_SOCmin}) guarantees the ability of $\widehat {SOC}$ to quickly provide a reliable  estimate of $SOC_{\min}$. 
Finally, in the next proposition, we ensure the existence of a dwell-time for any solution with bounded input, thereby ruling out the Zeno phenomenon.

\begin{prop}
        Given any $\ell>0$ and $\epsilon>0$, for any c\`adl\`ag input $u \in \mathcal{L}_\infty$, any solution $q$ to (\ref{compacthybridsystem}) has a dwell time, in particular there exists $M(q(0,0),\|u\|_\infty) \in \mathbb{R}_{>0}$ such that for any $(t,j)$, $(t',j') \in$ dom $q$ with $t+j<t'+j'$, $j'-j\leq \tfrac{1}{\tau}(t'-t)+1$ with $\tau:=\tfrac{\mu\varepsilon}{M(q(0,0),\|u\|_\infty)}$.
    $\hfill \square$
    \label{Prop_Zeno}
\end{prop}

The proof of Proposition \ref{Prop_Zeno} is omitted for space reasons. Proposition \ref{Prop_Zeno} implies that there exists a minimum amount of time between two successive jumps and hence rules out the presence of Zeno solutions. We see that $\varepsilon>0$ is needed to ensure the presence of the dwell-time and thus to eliminate the Zeno phenomenon.
\subsection{Proofs}
\label{StabilityGuar_Proofs}

\subsubsection{Proof of Proposition \ref{Th_StabilityProperty_delta_URC}}
Let $\mathcal{O}:=\mathbb{R}^{2N}\times \mathbb{R} \times \mathbb{R} \times \left\{\{1,\ldots,N\}+\iota [-1,1]^N\right\}$ with $ \iota\in(0,1)$.
We consider the Lyapunov function candidate $V_1(q):=\underset{i \in \{1,\hdots,N\}}{\max} (U_{RC,i}-\widehat{U}_{RC,i})^2$ for any $q \in \mathcal{O}$. We note that $V_1$ is locally Lipschitz  on an open set containing $\mathcal{Q}$. For any $q \in \mathcal{O}$, we have 
\begin{equation}
\frac{1}{N}  |U_{RC}-\widehat{U}_{RC}|^2 \leq V_1(q) \leq |U_{RC}-\widehat{U}_{RC}|^2.
\label{SandwishBound_on_Urc}
\end{equation}

Let $(q,u) \in \mathcal{C}$. By \cite[Lemma 1]{Liberzon_2012_CDC} and in view of the definition of $f$ below (\ref{compacthybridsystem}),
    $V_1^{\circ}(q;f(q,u)) \leq \underset{i \in  \{1,\hdots,N\}}{\max} \big\{ 2(U_{RC,i}-{\widehat{U}}_{RC,i})  \big(-\frac{1}{\tau_{d,i}}U_{RC,i}+\frac{1}{C_{d,i}}u-\frac{1}{C_{d,i}}\big(-\frac{1}{\tau_d}\overline{U}_{RC}+u\big)\big)\big\} \\
    \leq \underset{i \in  \{1,\hdots,N\}}{\max} \big\{ -\frac{2}{\tau_d} (U_{RC,i}-{\widehat{U}}_{RC,i})^2 +2\big(\frac{1}{\tau_d}-\frac{1}{\tau_{d,i}} \big) U_{RC,i} (U_{RC,i}-{\widehat{U}}_{RC,i})\big\}.$
Using the fact that for any $a,b \in \mathbb{R}$ and $\eta>0$, $2ab \leq \frac{\eta}{2} a^2 + \frac{2}{\eta} b^2$, we derive by taking $a=\left(\frac{1}{\tau_d}-\frac{1}{\tau_{d,i}} \right) U_{RC,i}$, $b=U_{RC,i}-{\widehat{U}}_{RC,i}$ and $\eta = 2 \tau_d$,
$  V_1^{\circ}(q;f(q,u))  \leq \underset{i \in  \{1,\hdots,N\}}{\max} \big\{ -\frac{2}{\tau_d} (U_{RC,i}-{\widehat{U}}_{RC,i})^2 +\tau_d\big(\frac{1}{\tau_d}-\frac{1}{\tau_{d,i}} \big)^2 {U_{RC,i}}^2 + \frac{1}{\tau_d} (U_{RC,i}-{\widehat{U}}_{RC,i})^2\big\}
   \leq  \underset{i \in  \{1,\hdots,N\}}{\max} \big\{ -\frac{1}{\tau_d} (U_{RC,i}-{\widehat{U}}_{RC,i})^2
   +\tau_d\big(\frac{1}{\tau_d}-\frac{1}{\tau_{d,i}} \big)^2 {U_{RC,i}}^2 \big\}.$

We thus have
\begin{equation}
\begin{aligned}
    V_1^{\circ}(q;f(q,u))& \leq - \tfrac{1}{\tau_d}V_1(q)  + \tau_d {d}^2 | U_{RC}|^2,
    \label{Flow_analysis_V1_2}
\end{aligned}
\end{equation}
with $d$ defined in Proposition \ref{Th_StabilityProperty_delta_URC}.

 Let $(q,u) \in \mathcal{D}$ and $g \in G(q)$, $V_1(g)=V_1(q)$ by (\ref{compacthybridsystem}).
 We then follow similar arguments as in the proof of Theorem 1 in \cite{petri2024hybrid} to obtain the desired result in (\ref{StabilityProperty_delta_URC}).

 \subsubsection{Proof of Proposition \ref{TH_Stability_Property_e}}
We consider the Lyapunov function candidate $V_2(q):={\max} \big\{  (SOC_{ \sigma }-\widehat{SOC})^2, \lambda V_1(q)\big\}$ for any $q \in \mathcal{O}$ with $\lambda:=\tfrac{4}{a_1^2}$ and $\mathcal{O}$ as in the proof of Proposition \ref{Th_StabilityProperty_delta_URC}. We note that $V_2$ is locally Lipschitz  on an open set containing $\mathcal{Q}$. For any $q \in \mathcal{O}$, we have 
\begin{equation}
   |SOC_\sigma-\widehat{SOC}|^2 \leq V_2(q) \leq b|e|^2,
    \label{sand_bound_v2}
\end{equation}
where $b=\max\{1,\lambda\}$.

Let $(q,u) \in \mathcal{C}$. To upper-bound $V_2^\circ (q; f(q,u))$, we apply \cite[Lemma 1]{Liberzon_2012_CDC} and distinguish the next three cases.

\textit{Case a):} $V_2(q)=(SOC_{ \sigma }-\widehat{SOC})^2 > \lambda  V_1(q)$. In view of the definition of $f$ below (\ref{compacthybridsystem}),  
$         V_2^{\circ}(q;f(q,u))= 2(SOC_{ \sigma }-\widehat{SOC})
        \left( -\tfrac{1}{3600Q_{ \sigma}}u-\left(-\tfrac{1}{3600Q_{ \sigma}}u+\ell(y_{ \sigma }-\hat{{y}})\right)\right)=-2\ell (SOC_{ \sigma }-\widehat{SOC})(y_{ \sigma }-\hat{{y}}).$
Given the expressions of $y_{ \sigma }$ and $\hat{y}$ after (\ref{compacthybridsystem}), $y_{ \sigma }-\hat{{y}}=V_{OCV}(SOC_{ \sigma})-V_{OCV}(\widehat{SOC})+\widehat{U}_{RC, \sigma}-{U}_{RC, \sigma}$, hence
\begin{equation}
    \begin{aligned}
        V_2^{\circ}(q;f(q,u))=&-2\ell (SOC_{ \sigma }-\widehat{SOC})\\&\quad \times (V_{OCV}(SOC_{ \sigma})-V_{OCV}(\widehat{SOC}))\\&+2\ell(SOC_{ \sigma }-\widehat{SOC})({U}_{RC, \sigma}-\widehat{U}_{RC, \sigma}).
    \end{aligned}
    \label{V2_flow_analysis_2}
\end{equation}
Since $V_{OCV}$ is continuously differentiable by item (i) of SA\ref{Ass_on_Vocv}, we derive by applying the mean value theorem that there exists $SOC' \in \mathbb{R}$ such that $V_{OCV}(SOC_{ \sigma})-V_{OCV}(\widehat{SOC})=\frac{\partial V_{OCV}}{\partial SOC}(SOC')(SOC_{ \sigma} -\widehat{SOC})$. Furthermore, given item (ii) of SA\ref{Ass_on_Vocv}, $(SOC_{ \sigma }-\widehat{SOC})(V_{OCV}(SOC_{ \sigma})-V_{OCV}(\widehat{SOC}))\geq a_1 (SOC_{ \sigma} -\widehat{SOC})^2$. On the other hand, as $(SOC_{ \sigma }-\widehat{SOC})^2 > \lambda  V_1(q)$, we get $|{U}_{RC, \sigma}-\widehat{U}_{RC, \sigma}| < \frac{1}{\sqrt{\lambda}}|SOC_{ \sigma }-\widehat{SOC}|$. Consequently, we derive from (\ref{V2_flow_analysis_2}) and the definition of $\lambda$
\begin{equation}
     V_2^{\circ}(q;f(q,u)) \leq -\ell a_1 V_2(q).
     \label{casea_Flow_analysis_V2}
\end{equation}

\textit{Case b):} $V_2(q)= \lambda V_1(q)  > (SOC_{ \sigma }-\widehat{SOC})^2 $. By (\ref{Flow_analysis_V1_2}), we have 
\begin{equation}
\begin{aligned}
    V_2^{\circ}(q;f(q,u))& \leq - \tfrac{1}{\tau_d}V_2(q)  + \lambda \tau_d {d}^2 | U_{RC}|^2.
    \label{caseb_Flow_analysis_V2}
\end{aligned}
\end{equation}

\textit{Case c):} $V_2(q)=(SOC_{ \sigma }-\widehat{SOC})^2= \lambda  V_1(q)$. Combining (\ref{casea_Flow_analysis_V2}) and (\ref{caseb_Flow_analysis_V2}), we obtain
\begin{equation}
\begin{aligned}
   V_2^{\circ}(q;f(q,u)) & \leq -aV_2(q)+\lambda \tau_d d^2 |U_{RC}|^2,
\end{aligned}
\label{casec_Flow_analysis_V2}
\end{equation}
where $a={\min}\big\{\ell a_1,\tfrac{1}{\tau_d}\big\}$.

In view of (\ref{casea_Flow_analysis_V2})-(\ref{casec_Flow_analysis_V2}), we have proved that for any $(q,u) \in \mathcal{C}$,
\begin{equation}
\begin{aligned}
   V_2^{\circ}(q;f(q,u)) & \leq -aV_2(q)+\lambda \tau_d d^2 |U_{RC}|^2.
\end{aligned}
\label{cas_Flow_analysis_V2}
\end{equation}

Let $(q,u) \in \mathcal{D}$ and $g \in G(q)$.
 We write for the sake of convenience $g=(x, { \overline{U}}_{RC}, {\widehat{SOC}}^+, \sigma^+)$ with some abuse of notation. We have seen in the proof of Proposition \ref{Th_StabilityProperty_delta_URC} that $V_1(g)=V_1(q)$, thus
 $    V_2(g) ={\max} \{  (SOC_{ \sigma^+ }-\widehat{SOC}^+)^2, \lambda V_1(q)\}.$
 We distinguish two cases below.
 
 \textit{Case a):} $V_2(g)=\lambda V_1(q) >  (SOC_{ \sigma^+ }-\widehat{SOC}^+)^2$. Then, 
 \begin{equation}
     \begin{aligned}
         V_2(g)\leq V_2(q).
     \end{aligned}
          \label{Jump_Analysis_casea_V2}
 \end{equation}
 
 \textit{Case b):} $V_2(g)= (SOC_{ \sigma^+ }-\widehat{SOC}^+)^2 > \lambda V_1(q)$. From the definition of $G$ below (\ref{compacthybridsystem}) and of $\mathcal{D}$ in (\ref{FlowJumpSets}), we have $V_{OCV}(\widehat{SOC}) - \varepsilon \leq  z_{ \sigma^+} \leq V_{OCV}(\widehat{SOC})-\mu \varepsilon$. This is equivalent to $V_{OCV}(\widehat{SOC}) - \varepsilon \leq V_{OCV}({SOC}_{ \sigma^+}) + \widehat{U}_{RC,\sigma^+}-U_{RC,\sigma^+} \leq V_{OCV}(\widehat{SOC})-\mu \varepsilon$. Therefore, as $\mu \in (0,1]$, we obtain $|V_{OCV}({SOC}_{ \sigma^+})-V_{OCV}(\widehat{SOC})| \leq |U_{RC,\sigma^+}-\widehat{U}_{RC,\sigma^+}|+\varepsilon$. Given item (i) in SA\ref{Ass_on_Vocv}, by the mean value theorem there exists $SOC' \in \mathbb{R}$ such that $V_{OCV}(SOC_{ \sigma^+})-V_{OCV}(\widehat{SOC})=\frac{\partial V_{OCV}}{\partial SOC}(SOC')(SOC_{ \sigma^+} -\widehat{SOC})$. Furthermore, given item (ii) of SA\ref{Ass_on_Vocv}, $|V_{OCV}(SOC_{ \sigma^+})-V_{OCV}(\widehat{SOC})| \geq a_1 |SOC_{ \sigma^+} -\widehat{SOC}|$. Thus, we obtain $  a_1 |SOC_{ \sigma^+} -\widehat{SOC}| \leq |U_{RC,\sigma^+}-\widehat{U}_{RC,\sigma^+}|+\varepsilon$. As $(U_{RC,\sigma^+}-\widehat{U}_{RC,\sigma^+})^2 \leq V_1(q)$, we get 
 \begin{equation}
     \begin{aligned}
 a_1 |SOC_{ \sigma^+} -\widehat{SOC}| \leq \sqrt{V_1(q)}+\varepsilon \leq \tfrac{1}{\sqrt{\lambda}}\sqrt{V_2(q)}+\varepsilon.
 \end{aligned}
      \label{V2_D_b_1}
  \end{equation}
On the other hand, from the definition of $G$ below (\ref{compacthybridsystem}) and of $\mathcal{D}$ in (\ref{FlowJumpSets}), we have $V_{OCV}(\widehat{SOC}) - \varepsilon \leq  \underset{i \in \{1,\hdots,N\}\backslash\{\sigma\}}{\min} z_i\leq V_{OCV}(\widehat{SOC})-\mu \varepsilon$. This is equivalent to $V_{OCV}(\widehat{SOC}) - \varepsilon \leq V_{OCV}(\widehat{SOC}^+) \leq V_{OCV}(\widehat{SOC})-\mu \varepsilon$. As $\mu \in (0,1]$, we obtain $|V_{OCV}(\widehat{SOC})-V_{OCV}(\widehat{SOC}^+)| \leq \varepsilon$. Given item (i) in SA\ref{Ass_on_Vocv}, by the mean value theorem there exists $SOC' \in \mathbb{R}$ such that $V_{OCV}(\widehat{SOC})-V_{OCV}(\widehat{SOC}^+)=\frac{\partial V_{OCV}}{\partial SOC}(SOC')(\widehat{SOC} -\widehat{SOC}^+)$. Furthermore, given item (ii) of SA\ref{Ass_on_Vocv}, $|V_{OCV}(\widehat{SOC})-V_{OCV}(\widehat{SOC}^+)| \geq a_1 |\widehat{SOC} -\widehat{SOC}|$. Thus, we obtain
$ a_1 |\widehat{SOC} -\widehat{SOC}^+| \leq \varepsilon.$
From the last inequality and (\ref{V2_D_b_1}),
since $|SOC_{\sigma^+}-\widehat{SOC}^+| \leq |SOC_{\sigma^+}-\widehat{SOC}|+|\widehat{SOC}-\widehat{SOC}^+|$, we derive by using $(a+b)^2 \leq 2a^2+2b^2$ for any $a,b \in \mathbb{R}$ and the definition of $\lambda$,
\begin{equation}
    \begin{aligned}
         V_2(g) \leq  \tfrac{1}{2}V_2(q)+\tfrac{8\varepsilon^2}{a_1^2}.
    \end{aligned}
        \label{Jump_Analysis_V2}
\end{equation}
Let $u$ be c\`adl\`ag and $q$ be a solution to system (\ref{compacthybridsystem}). Pick any $(t,j) \in $ dom $q$ and let $0 =t_0 \leq t_1 \leq \hdots \leq t_{j+1}= t$ satisfy dom $q \cap ([0,t] \times \{0, 1 \hdots j\})$.
For each $i \in \{0, \hdots, j\}$ and all $s \in [t_i,t_{i+1}]$, $(q(s,i),u(s)) \in \mathcal{C}$. Let $i\in \{0, \hdots, j\}$. In view of (\ref{cas_Flow_analysis_V2}), for almost all $s\in[t_i,t_{i+1}]$,
$         V_2^{\circ}(q(s,i);f(q(s,i),u(s,i)))\leq - a V_2(q(s,i)) + \lambda \tau_d d^2 \|U_{RC}\|^2_{(s,i)}.$
In view of \cite{Teel_Praly_MCSS_2000}, for almost all $s\in[t_i,t_{i+1}]$,
$    \frac{d}{ds}V_2(q(s,i)) \leq V_2^{\circ}(q(s,i);f(q(s,i),u(s,i))).$
Hence, for almost all $s\in[t_i,t_{i+1}]$,
$     \frac{d}{ds}V_2(q(s,i)) \leq - a V_2(q(s,i)) + \lambda \tau_d d^2 \|U_{RC}\|^2_{(s,i)}.$
Applying the comparison principle \cite[Lemma 3.4]{khalil_2002_nonlinearsystemsbook}, we obtain for all $s\in[t_i,t_{i+1}]$, 
\begin{equation}
    \begin{aligned}
         V_2(q(s,i))  \leq & V_2(q(t_i,i))e^{-a(s-t_i)}\\&+\tfrac{1}{a}(1-e^{-a(s-t_i)})\lambda \tau_d d^2 \|U_{RC}\|^2_{(s,i)}.
    \end{aligned}
        \label{flow_analysis_V2_3}
\end{equation}

On the other hand, for each $i \in \{1, \hdots, j\}$, $(q(t_i,i-1),u(t_i)) \in \mathcal{D}$. From (\ref{Jump_Analysis_casea_V2}) and (\ref{Jump_Analysis_V2}), we obtain for any $i \in \{1, \hdots, j\}$
\begin{equation}
    V_2(q(t_i,i)) \leq \max\{V_2(q(t_i,i-1)), \tfrac{1}{2}V_2(q(t_i,i-1))+\tfrac{8\varepsilon^2}{a_1^2}\}.
    \label{Jump_analysis_V2_2}
\end{equation}
From (\ref{flow_analysis_V2_3}) and (\ref{Jump_analysis_V2_2}), we deduce for any $(t,j) \in $ dom $q$
\begin{equation}
    \begin{aligned}
         V_2(q(t,j)) \leq&  V_2(q(0,0))e^{-at}
         + \tfrac{8\varepsilon^2}{a_1^2}\sum_{i=0}^{j-1} \left(\tfrac{1}{2}\right)^ie^{-a(t-t_{j-i})} 
         \\& + (1-e^{-at}) \tfrac{\lambda {\tau_d}}{a}d^2 \|U_{RC}\|^2_{(t,j)}.
    \end{aligned}
    \label{V2_bound}
\end{equation}
We also have $\sum_{i=0}^{j-1} \left(\tfrac{1}{2}\right)^ie^{-a(t-t_{j-i})} \leq \sum_{i=0}^{j-1} \left(\tfrac{1}{2}\right)^i\leq 2.$ 
Using (\ref{sand_bound_v2}) in (\ref{V2_bound}) and by using $\sqrt{x+y} \leq \sqrt{x}+\sqrt{y}$ for all $x ,y \in \mathbb{R}_{\geq0} $, we obtain (\ref{Th1_property}). 

\subsubsection{Proof of Theorem \ref{TH_Stability_Property_SOCmin}}
 Let $u$ be c\`adl\`ag and $q$ be a corresponding solution to (\ref{compacthybridsystem}). 
 In view of the definitions of $\mathcal{C}$ and $\mathcal{D}$ in (\ref{FlowJumpSets}), for any $(t,j) \in$ dom $q$ and any $i \in \{1,\hdots,N\}\backslash\{{ \sigma(t,j)}\}$, $V_{OCV}(\widehat{{SOC}}(t,j))\leq z_i(t,j) + \varepsilon$. From the definition of $z_i$ in (\ref{zi}), we obtain $V_{OCV}(\widehat{{SOC}}(t,j)) \leq V_{OCV}(SOC_i(t,j))-U_{RC,i}(t,j)+\widehat{U}_{RC,i}(t,j)+\varepsilon$. 
 This implies for any $(t,j) \in$ dom $q$,
 \begin{equation}
     \begin{aligned}
         V_{OCV}(\widehat{{SOC}}&(t,j)) - V_{OCV}(SOC_{\min}(t,j))\\&\leq \widehat{U}_{RC,m(t,j)}(t,j)-U_{RC,m(t,j)}(t,j)+\varepsilon,
     \end{aligned}
 \end{equation}
 where $m(t,j) \in \underset{i \in \{1,\hdots,N\}\backslash\{{ \sigma}\}}{\argmin}V_{OCV}(SOC_i(t,j))$.

Let $(t,j) \in$ dom $q$, we next distinguish three cases.

\textit{Case a):}  $SOC_{\min}(t,j)=SOC_{\sigma(t,j)}(t,j)$. Since the conditions of Proposition \ref{TH_Stability_Property_e} are verified, (\ref{Th1_property}) holds, hence, 
\begin{equation}
    \begin{aligned}
        |SOC_{\min}(t,j)-\widehat{SOC}(t,j)|\leq & c_1|e(0,0)|e^{-c_2t} \\&+ c_3 {\varepsilon}+c_4 d {\|U_{RC}\|_{(t,j)}}.
    \end{aligned}
    \label{casea_SOCmin_Prop}
\end{equation}

\textit{Case b):} $SOC_{\min}(t,j)\leq \widehat{SOC}(t,j)$ \textit{with} $SOC_{\min}(t,j) \neq SOC_{\sigma}(t,j)$. By item (i) of SA\ref{Ass_on_Vocv}, we obtain by applying the mean value theorem that there exists $SOC'$ in $(SOC_{\min}(t,j),\widehat{{SOC}}(t,j))$ such that $V_{OCV}(\widehat{{SOC}}(t,j))-V_{OCV}({{SOC}}_{\min}(t,j))=\frac{\partial V_{OCV}}{\partial SOC}(SOC')(\widehat{{SOC}}(t,j) -SOC_{\min}(t,j))$. Hence, as $SOC_{\min}(t,j)\leq \widehat{SOC}(t,j)$, by item (ii) of SA\ref{Ass_on_Vocv}, $a_1 (\widehat{{SOC}}(t,j) -SOC_{\min}(t,j))\leq \widehat{U}_{RC,m{(t,j)}}(t,j)-U_{RC,m{(t,j)}}(t,j)+ \varepsilon$. Equivalently, 
  \begin{equation}
  \begin{aligned}
      |\widehat{{SOC}}(t,j)& -SOC_{\min} (t,j)|\\&\leq \tfrac{1}{a_1}(\widehat{U}_{RC,m{(t,j)}}(t,j)-U_{RC,m{(t,j)}}(t,j))+\tfrac{\varepsilon}{a_1}.
      \label{deltaSOC_property}
      \end{aligned}
  \end{equation}
Given that the conditions of Proposition \ref{Th_StabilityProperty_delta_URC} are verified, $q$ satisfies (\ref{StabilityProperty_delta_URC}). Hence, from (\ref{StabilityProperty_delta_URC}) and (\ref{deltaSOC_property}), we derive  
\begin{equation}
    \begin{aligned}
        |\widehat{{SOC}}(t,j)& -SOC_{\min}(t,j)|  \\&\leq \tfrac{\sqrt{N}}{a_1} |U_{RC}(0,0)-\widehat{U}_{RC}(0,0)| e^{-\frac{1}{2\tau_d}t} \\&+ \tfrac{\sqrt{N}\tau_d }{a_1} d \|U_{RC}\|_{(t,j)} + \tfrac{\varepsilon}{a_1}.
    \end{aligned}
        \label{caseb_SOCmin_Prop}
\end{equation}

\textit{Case c):} $SOC_{\min}(t,j)>\widehat{SOC}(t,j)$. We have 
$SOC_{\min}(t,j)-\widehat{SOC}(t,j)=SOC_{\min}(t,j)-SOC_{\sigma(t,j)}(t,j)+SOC_{\sigma(t,j)}(t,j)-\widehat{SOC}(t,j)$. As $SOC_{\min}(t,j) \leq SOC_{\sigma(t,j)}(t,j)$ by definition in (\ref{Def_LimitingSOC}), we derive $   |SOC_{\min}(t,j)-\widehat{SOC}(t,j)|\leq  SOC_{\sigma(t,j)}(t,j)-\widehat{SOC}(t,j).$
Since the conditions of Proposition \ref{TH_Stability_Property_e} are verified, (\ref{Th1_property}) holds, we thus obtain 
\begin{equation}
    \begin{aligned}
        |SOC_{\min}(t,j)-\widehat{SOC}(t,j)|\leq & c_1|e(0,0)|e^{-c_2t} \\&+ c_3 {\varepsilon}+c_4 d {\|U_{RC}\|_{(t,j)}}.
    \end{aligned}
    \label{casec_SOCmin_Prop}
\end{equation}
From (\ref{casea_SOCmin_Prop}), (\ref{caseb_SOCmin_Prop}) and (\ref{casec_SOCmin_Prop}), we derive the desired stability property in (\ref{StabilityProperty_with_SOCmin}).
\section{Numerical illustration}
\label{Simu}
\begin{figure}[h]
    \centering
\includegraphics[scale=0.5]{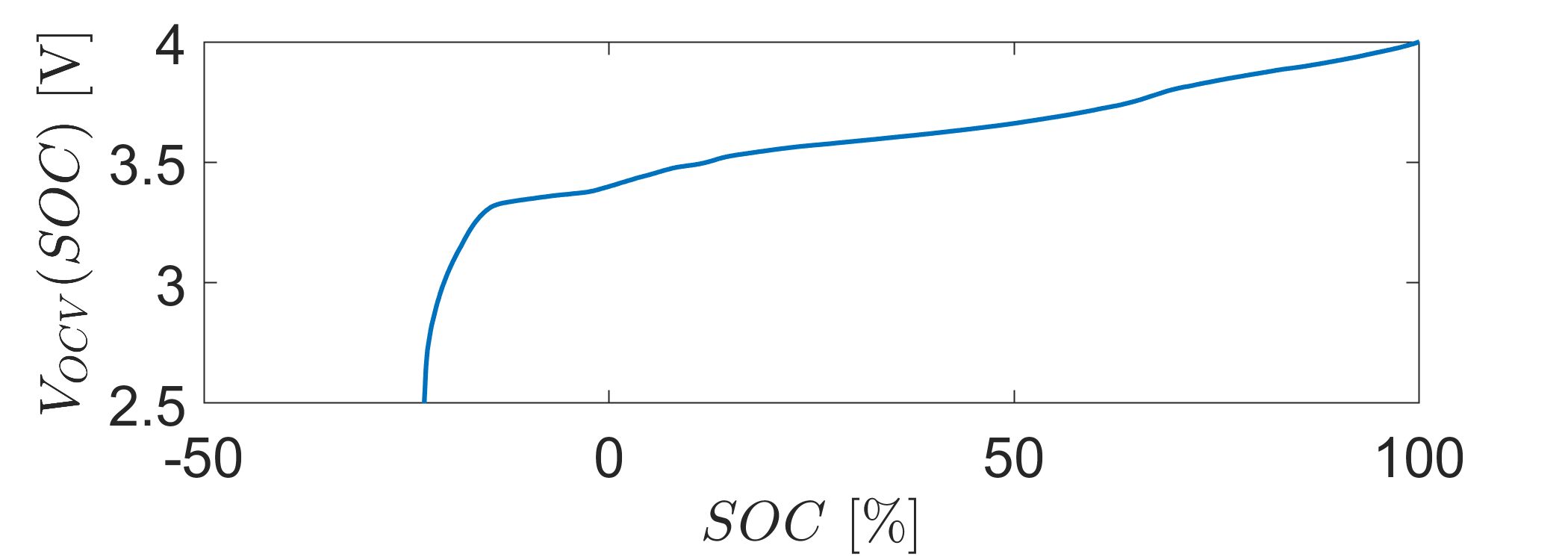}
\vspace{-0.7cm}
    \caption{$V_{OCV}$ function.}\vspace{-0.4cm}
    \label{Fig_OCVSOCmap}
\end{figure}

\begin{figure}[h]
    \centering
\includegraphics[scale=0.5]{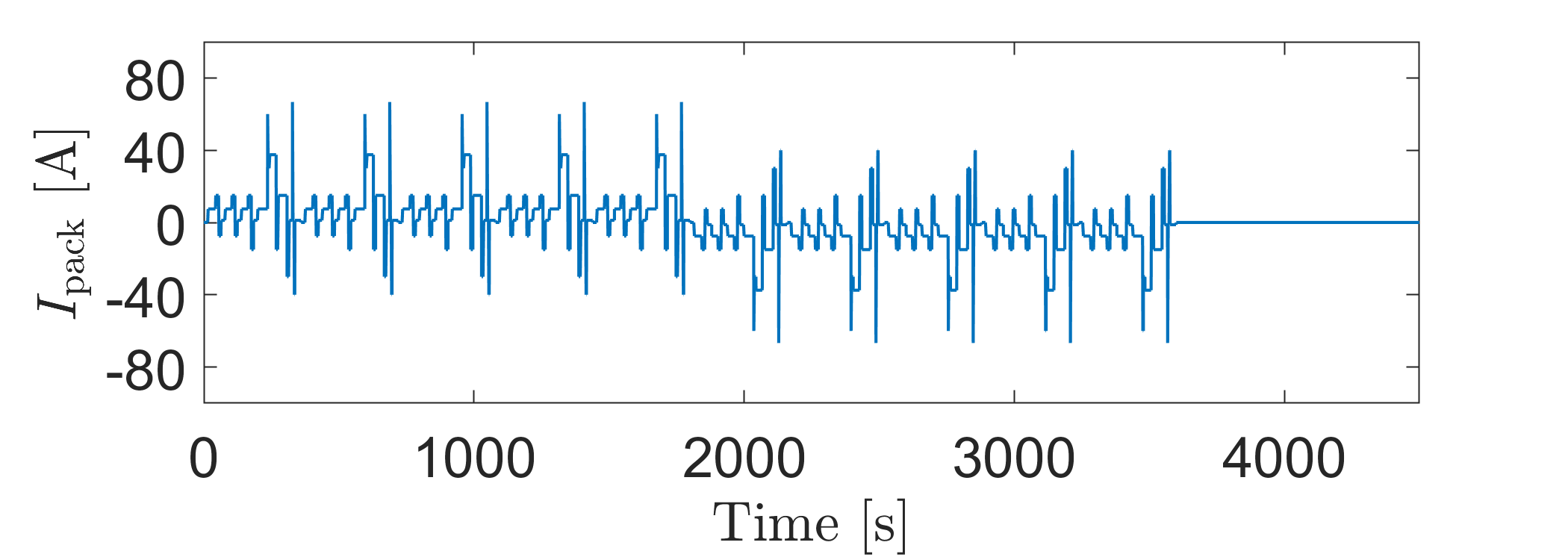}
\vspace{-0.7cm}
    \caption{Input current profile.}\vspace{-0.5cm}
    \label{Fig_Ipack}
\end{figure}

We illustrate the effectiveness of the hybrid estimator. 
We consider for this purpose a battery pack consisting of $200$ cells interconnected in series. The cells parameters are taken from a normal distribution with $10\%$ dispersion from the nominal parameters $\tau_{d,\text{nom}}=12$ s, $R_{d,\text{nom}}=0.5$ m$\Omega$, $R_{\text{int},\text{nom}}=0.5$ m$\Omega$ and $Q_{\text{nom}}=6$ Ah. We initialized the SOC of the cells slightly differently to create additional disparities between the cells. The considered $V_{OCV}$ function for all the cells is depicted in Figure \ref{Fig_OCVSOCmap}, it corresponds to a graphite negative electrode and a NCA positive electrode technology and thus verifies SA\ref{Ass_on_Vocv} with $a_1=2.3$ mV/\% and $a_2=616.6$ mV/\%. 
The input current $u$ is  a
plug-in hybrid electrical vehicle (PHEV) current profile, see Figure \ref{Fig_Ipack}. 


We have simulated the hybrid model in (\ref{compacthybridsystem}) by taking $\ell=2$, $\tau_d=12$ s, $\varepsilon=10^{-3}$ and $\mu=0.95$. We have initialized the hybrid estimator with $\sigma(0,0)=150$, $\widehat{SOC}(0,0)=0$ and $\overline{U}_{RC}(0,0)=0$. We note that $q(0,0)$ is in $\mathcal{C}\cup\mathcal{D}$ as discussed in Section \ref{ Hybriddesign_FlowandJumpsets} and $\sigma(0,0)$ does not correspond to the index of the cell with $\widehat{SOC}_{\min}(0,0)$.
\begin{figure}[h]
    \centering
\includegraphics[scale=0.5]{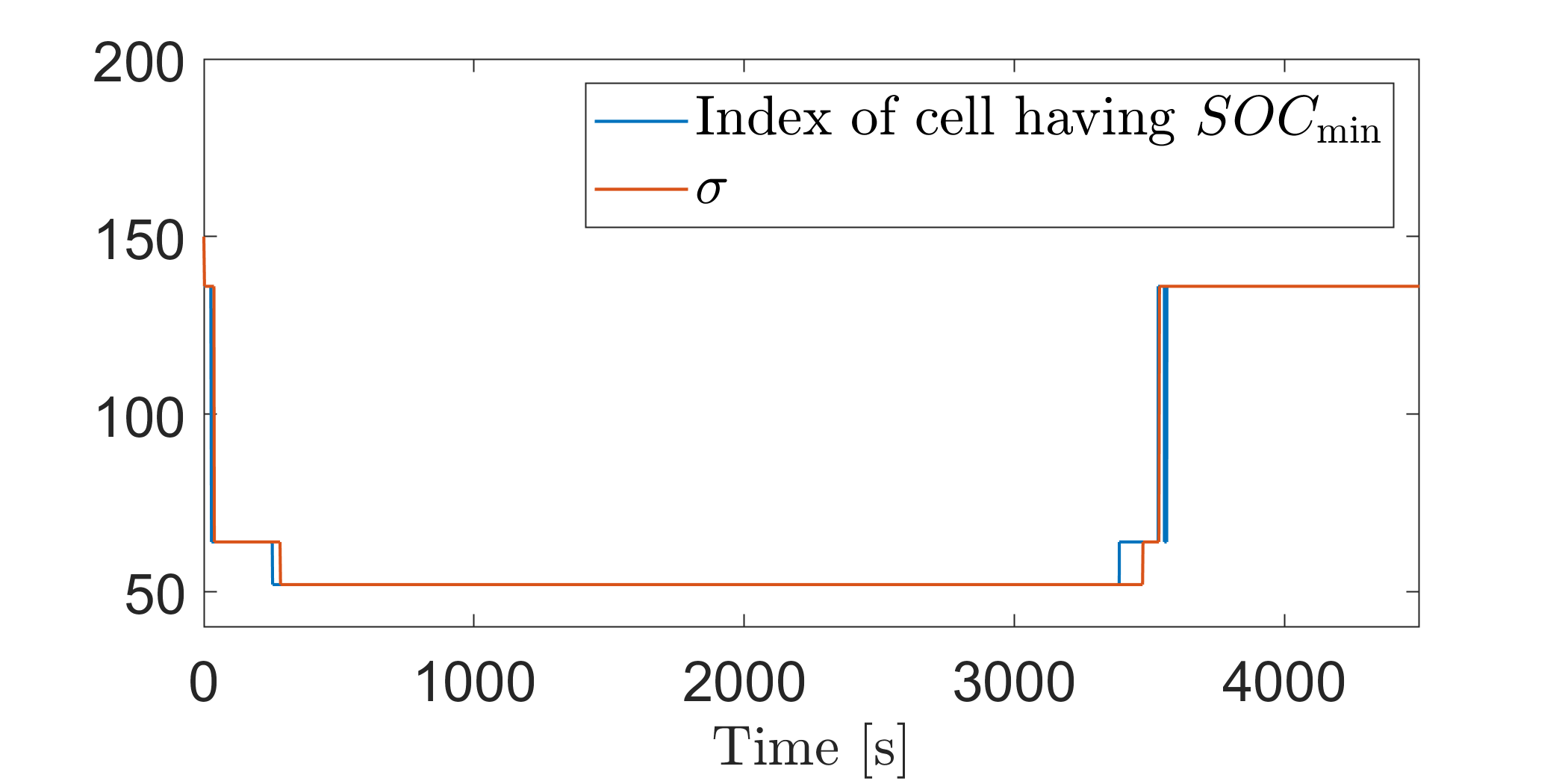}
\vspace{-0.7cm}
    \caption{The index of the cell having $SOC_{\min}$ and $\sigma$.}\vspace{-0.4cm}
    \label{Fig_cellindexmin}
\end{figure}

\begin{figure}[h]
    \centering
\includegraphics[scale=0.5]{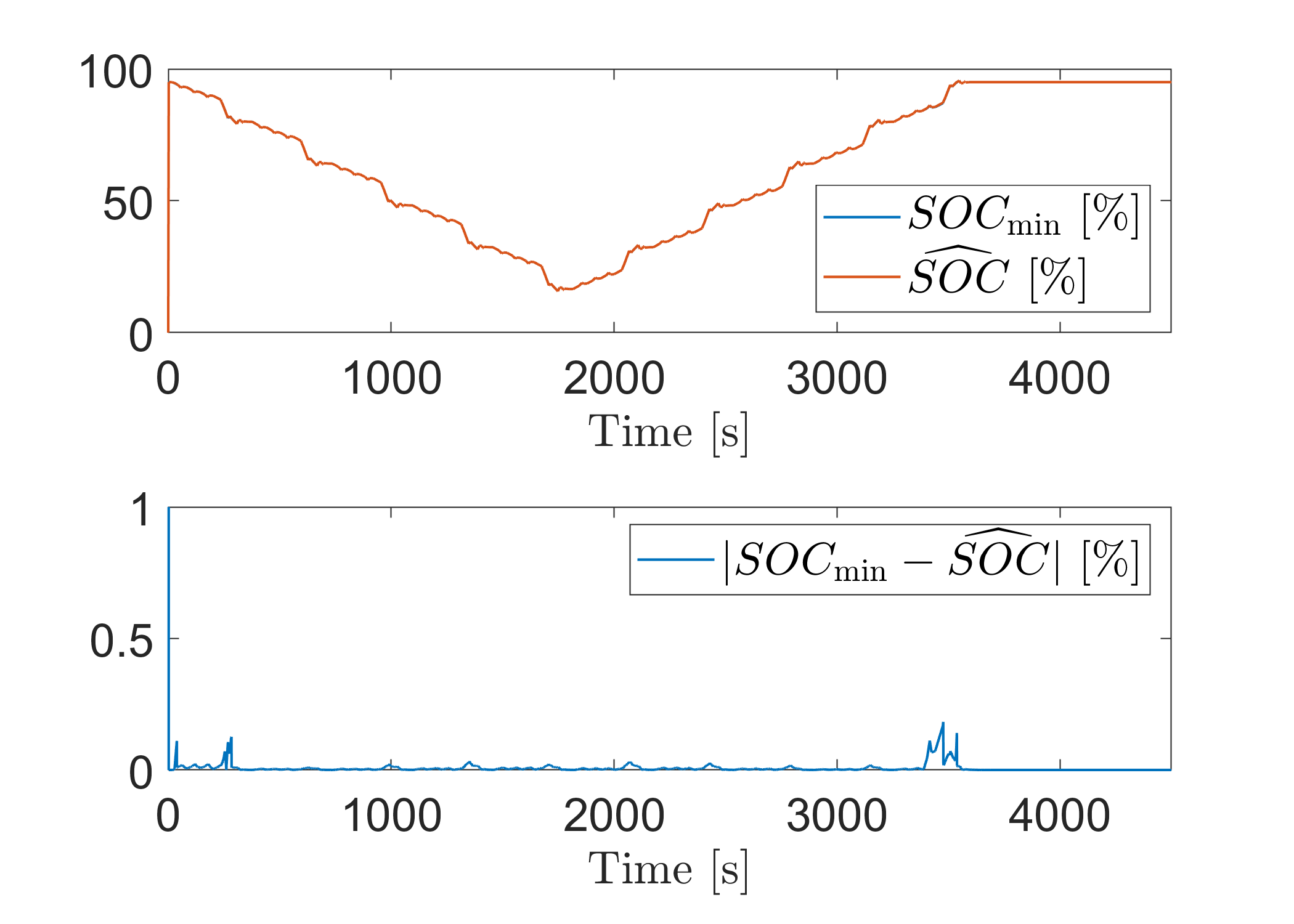}
\vspace{-0.7cm}
    \caption{$SOC_{\min}$ et $\widehat{SOC}$ generated by the hybrid estimator (top) and the norm of the estimation error $|SOC_{\min}-\widehat{SOC}|$ (bottom).}\vspace{-0.5cm}
    \label{Fig_SOCminetSocminEst}
\end{figure}
Figure \ref{Fig_cellindexmin} shows the index of the cell having the minimum SOC over time and $\sigma$ the index of the cell based on which $\widehat{SOC}$ is generated. We see that $\sigma$ corresponds to the cell with the minimum SOC most of the time, but not always. Still, Figure \ref{Fig_SOCminetSocminEst}, which reports the minimum SOC and the estimated minimum SOC generated by the hybrid estimator, as well as the
corresponding norm of the estimation error on the minimum SOC, shows that the hybrid estimator indeed quickly provides a reliable estimate of the minimum SOC.  

\section{Conclusion}
We have designed a low-dimensional hybrid estimator for the minimum SOC of a battery pack consisting of cells interconnected in series, each being modeled by a first order ECM. A practical exponential stability property for the minimum SOC estimation error was established. The obtained simulation results illustrate the relevance of the proposed hybrid estimation scheme. In future work, we will explain how to \textit{simultaneously} estimate the minimum and the maximum SOC and we also plan to take into account measurement noises in the analysis.
\label{conclu}

\addtolength{\textheight}{-12cm}   





\bibliography{IEEEabrv,paper.bib}

\end{document}